\documentclass[sigconf,screen]{acmart}
\usepackage{graphicx} % Required for inserting images
\usepackage{array}
\usepackage{threeparttable}
\usepackage{enumitem}
\usepackage{color}
\usepackage{tcolorbox}
\usepackage[table]{xcolor}
\usepackage{multirow}
\usepackage{ulem}
\usepackage{pifont}
\usepackage[linesnumbered,ruled,vlined]{algorithm2e}
\usepackage{amsmath} 
\usepackage[T1]{fontenc}
\usepackage{svg}
\usepackage{listings}
\usepackage{tikz}
\usetikzlibrary{arrows.meta,positioning,fit}
\usepackage{tabularx}
\usepackage{tcolorbox}
\tcbuselibrary{listings, skins, breakable}

\setcopyright{cc}
\setcctype{by}
\acmDOI{10.1145/3844178.3844646}
\acmYear{2026}
\copyrightyear{2026}
\acmISBN{979-8-4007-3002-3/2026/10}
\acmConference[SWGeno '26]{Proceedings of the 2nd International Workshop on Software Genomics}{October 12--16, 2026}{Munich, Germany}
\acmBooktitle{Proceedings of the 2nd International Workshop on Software Genomics (SWGeno '26), October 12--16, 2026, Munich, Germany}
\acmSubmissionID{asews26swgenomain-p59-p}
\received{2026-07-25}
\received[accepted]{2026-08-21}

\newtcblisting{ccode}{
    listing only,
    listing options={
        language=C,
        basicstyle=\ttfamily\small,
        breaklines=true,
        breakatwhitespace=false,
        columns=flexible,
        escapechar=@ 
    },
    colback=gray!5,
    colframe=gray!75,
    arc=2mm,
    boxrule=0.5pt,
    left=0mm,
    top=0mm,
    bottom=0mm,
    breakable
}

\newtcblisting{pycode}{
    listing only,
    listing options={
        language=Python,
        basicstyle=\ttfamily\small,
        breaklines=true,
        breakatwhitespace=false,
        columns=flexible,
        escapechar=@
    },
    colback=gray!5,
    colframe=gray!75,
    arc=2mm,
    boxrule=0.5pt,
    left=0mm,
    top=0mm,
    bottom=0mm,
    breakable
}

\usepackage[commandnameprefix=ifneeded]{changes}
\setauthormarkup{}
\definechangesauthor[name={ttfish}, color=olive]{ttfish}

\title{The Gene Is Not the Gate: Conceptualizing the Semantic Genome of Quantum Software}

\author{Jiongchi Yu}
\orcid{0000-0002-2888-4499}
\affiliation{%
  \institution{Nanyang Technological University}
  \city{Singapore}
  \country{Singapore}
}
\email{jiongchiyu@acm.org}

\author{Yue Duan}
\orcid{0000-0003-1049-9645}
\affiliation{%
  \institution{Singapore Management University}
  \city{Singapore}
  \country{Singapore}
}
\email{yueduan@smu.edu.sg}

\author{Ziming Zhao}
\orcid{0000-0003-1455-4330}
\affiliation{%
  \institution{Zhejiang University}
  \city{Hangzhou}
  \country{China}
}
\email{zhaoziming@zju.edu.cn}

\author{Xiaofei Yue}
\orcid{0009-0005-2482-579X}
\affiliation{%
  \institution{Beijing Institute of Technology}
  \city{Beijing}
  \country{China}
}
\email{xfyue@bit.edu.cn}

\author{Zhaoxuan Li}
\orcid{0000-0002-2195-0799}
\affiliation{%
  \institution{Institute of Information Engineering, CAS \& University of Chinese Academy of Sciences}
  \city{Beijing}
  \country{China}
}
\email{lizhaoxuan@iie.ac.cn}



\begin{document}

% \keywords{Static analysis, Dynamic validation, Vulnerability detection, Proof of Concept (PoC), Large models}

\begin{abstract}
Quantum software spans applications, program representations, compilers, runtimes, and hardware-facing control. Existing abstractions describe these artifacts and layers, but they do not capture the semantic responsibilities that must remain valid as a computation is transformed across the stack. This limits systematic reasoning about cross-layer correctness, evolution, and reuse.

We present \textit{QGenome}, a semantic foundation for quantum software genomics. We define a quantum software gene as a typed, reusable responsibility characterized by an explicit contract, execution context, concrete expressions, and supporting evidence. QGenome organizes genes across five software strata and ten responsibility families grouped into computational semantics, coordination, realization, and assurance. Typed genes compose into pathways and genomes through dependency, refinement, provenance, and phenotype relations, enabling responsibilities to be traced across alternative implementations and execution contexts.

We illustrate the taxonomy using representative public artifacts across the quantum software stack and define principles for contextual validity, genome evolution, environmental adaptation, and capability expression. Rather than presenting a complete taxonomy or an automated extraction technique, QGenome establishes a falsifiable framework and a concrete research agenda for gene mining, lineage recovery, cross-layer regression testing and repair, and evidence-grounded quantum software agents.

% Quantum software includes applications, languages, compilers, hybrid
% runtimes, resource managers, controllers, calibration services, simulators,
% and assurance tools.  Existing representations identify where software
% resides or how an artifact is encoded, but not the internal responsibility
% that should remain valid when a computation is transformed across strata.
% We argue that a quantum software gene is a typed, contract-bearing semantic
% responsibility rather than a gate sequence.  QGenome classifies each gene
% along two independent axes: five software strata and ten responsibility
% families grouped under computational semantics, coordination, realization,
% and assurance.  A genome is an attributed graph of genes, dependencies,
% alternative expressions, refinement links, contexts, and evidence.  To ground
% the proposal, we map reported features of 14 public systems---including
% % ProjectQ, OpenQL, XACC/qcor, Tierkreis, OpenQASM~3, Pulser, Qibolab, MQT
% Bench, LintQ, MorphQ, and Bugs4Q---to the taxonomy and release a
% machine-checkable evidence catalog.  We define construction, composition,
% evolution, environmental adaptation, and capability-expression obligations,
% and state relative completeness as a falsifiable coverage criterion rather
% than an accomplished universal result.  This foundation connects reuse,
% assurance, and change analysis across the full quantum software ecosystem.

\end{abstract}

\maketitle

\section{Introduction}

Quantum software spans a layered ecosystem from application abstractions to hardware-facing control. Circuits, source programs, intermediate representations, runtimes, and control systems expose different stages of this stack~\cite{steiger2018projectq,khammassi2022openql,mccaskey2020xacc,mccaskey2021qcor,sivarajah2022tierkreis,cross2022openqasm,qir}. These abstractions describe where a computation is represented or transformed, but they do not identify the finer-grained semantic responsibilities that must remain valid across layers.

\begin{figure*}[t]
    \centering
    \includegraphics[width=.9\linewidth]{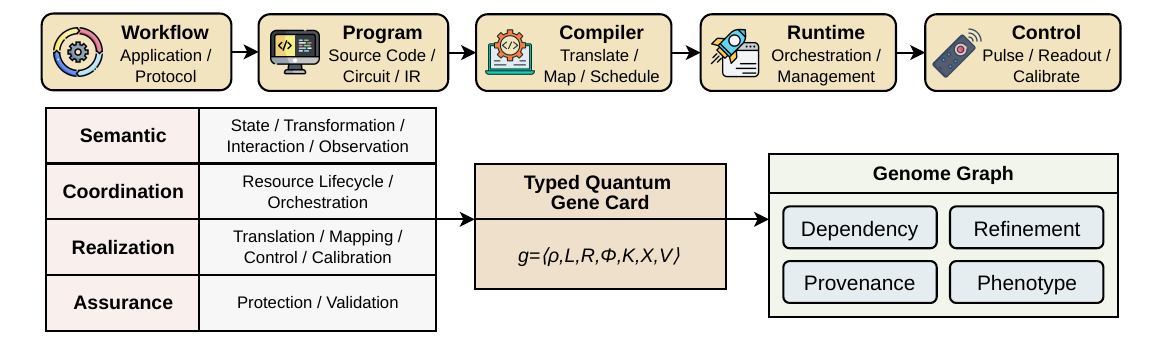}
    \caption{QGenome separates software strata from responsibility families.
    A typed gene card connects both dimensions through an explicit semantic
    contract, while genome graphs relate gene cards through dependency,
    refinement, provenance, and phenotype links.}
    \label{fig:qgenome}
\end{figure*}

Bell-state preparation illustrates this gap. Its meaning lies in the state relation established between two logical qubits rather than in a specific realization such as \textit{H} followed by \textit{CX}. A compiler may decompose or reroute these operations as long as the intended relation is preserved. The same requirement becomes more demanding in teleportation, where measurement outcomes must control corrections on another qubit. A simulator may execute this workflow, while a target backend may lack the required control capability or timing support. Calibration changes can further alter the physical realization without modifying the source program or its intermediate representation. These cases are therefore different expressions of related semantic responsibilities rather than isolated artifacts.

Existing abstractions do not expose these relationships directly. Gate motifs capture local implementations, while stack models record transformation locations. Neither states what must be preserved or what evidence supports that preservation. This limitation also extends beyond gate-model programs. Tierkreis, Pulser, and Qibolab expose workflow, pulse-control, and hardware-facing responsibilities that cannot be reduced to a single notion of a quantum program~\cite{sivarajah2022tierkreis,silverio2022pulser,efthymiou2024qibolab}. Even measurement serves different purposes in application readout, syndrome extraction, and controller discrimination. The appropriate unit of analysis must therefore be finer-grained than a software product, richer than a syntactic pattern, and explicitly tied to both semantic role and software stratum. This need is consistent with evidence that quantum software faults often emerge from interactions among source code, circuits, qubit usage, and platform behavior~\cite{paltenghi2022bugs,paltenghi2024lintq}.

\textbf{Position.} We define a quantum software gene as a typed, reusable semantic responsibility with a contract that remains checkable across the strata required for its realization. Each gene records its semantic subject, layer span, execution context, observable expressions, and supporting evidence, covering applications, compilers, runtimes, controllers, workflows, simulators, and assurance tools. We introduce \textit{QGenome}, as shown in \autoref{fig:qgenome}, which organizes these responsibilities as typed gene cards across software strata and responsibility families. Contracts, dependencies, refinement, provenance, and cross-layer expression relations compose genes into pathways and genomes. \textit{QGenome} is an extensible foundation, not a complete taxonomy or automatic extraction method.
\section{Conceptual Foundations} 

\textbf{Quantum Software Genes.} We define a quantum software gene as a reusable semantic responsibility whose realization and preservation can be traced across the relevant software strata. A gene associates this responsibility with the software object that owns or enforces it, an explicit semantic contract, and the context and evidence needed to justify its realization. Here, ownership refers to the program element, compiler pass, runtime service, control module, or other software object responsible for satisfying the contract. Unlike a design pattern, intermediate representation, component, or test, a gene connects these artifacts through a shared and checkable semantic obligation.

We represent a gene card as $g=\langle \rho,O,L,F,R,\Phi,K,X,V\rangle$. Here, $\rho$ captures the semantic responsibility and contract subject, $O$ its owner, $L$ its layer span, $F$ its taxonomy labels, and $R$ its resource roles. The contract $\Phi=(A,G)$ specifies the semantic assumptions $A$ and guarantees $G$ of the responsibility. The context $K$ records realization-specific conditions, while $X$ denotes the concrete expressions and $V$ provides evidence that $K$ satisfies $A$ and that $X$ realizes $G$ under $K$. Gene identity is determined by $\rho$, $R$, and $\Phi$, while versions and lineage are maintained in the genome graph.

\textbf{Scope of Classification.} Our taxonomy classifies documented quantum-dependent responsibilities rather than entire repositories. Each candidate must specify a contract subject, an owning or enforcing software object, at least one observable expression, and public evidence. Candidates with separable obligations over different subjects are split into gene cards linked by dependency or refinement. Multiple family labels are used only for inseparable cross-family obligations. Insufficiently evidenced candidates remain unresolved rather than established genes.

% This design makes the taxonomy reproducible and falsifiable, allowing researchers to inspect individual assignments and refine the classification as new responsibilities emerge.
\section{Quantum Software Gene Taxonomy}

QGenome classifies each gene along two orthogonal axes. The \textit{software-stratum axis} identifies where a responsibility is realized, while the \textit{responsibility-family axis} specifies the semantic obligation it governs. This separation is necessary because one responsibility may span multiple strata, while one software system may own several kinds of genes~\cite{haener2018methodology,mccaskey2020xacc}. As summarized in Table~\ref{tab:taxonomy}, the taxonomy contains ten responsibility families organized into four groups: computational semantics, coordination, realization, and assurance.

\textbf{Computational Semantics.} These genes govern how quantum information is prepared, transformed, related, and observed. \textit{State}, \textit{transformation}, \textit{interaction}, and \textit{observation} genes capture these respective responsibilities. They are reusable across algorithms: teleportation composes state and observation genes with classically controlled correction, while VQE combines state and transformation genes with observation in a classical optimization loop.

\textbf{Coordination.} These genes organize resources and hybrid execution. \textit{Resource-lifecycle genes} govern allocation, reset, reuse, and release, whereas \textit{orchestration genes} capture dependencies between quantum execution and classical control. Tierkreis realizes orchestration through typed quantum-classical workflows, while XACC/qcor coordinates quantum kernels, accelerator selection, and hybrid execution ~\cite{sivarajah2022tierkreis,mccaskey2020xacc,mccaskey2021qcor}.

\textbf{Realization.} These genes connect abstract computations to executable device operations. \textit{Translation genes} preserve meaning across source, IR, and native representations. \textit{Mapping-and-timing genes} bind logical operations to physical resources and schedules. \textit{Control-and-calibration genes} connect scheduled operations to signals and device parameters. OpenQL exposes translation and mapping responsibilities, while Pulser and Qibolab provide different forms of pulse-level and hardware-facing control ~\cite{khammassi2022openql,silverio2022pulser,efthymiou2024qibolab}.

\textbf{Assurance.} These genes establish or preserve claims about correctness and execution quality. \textit{Protection-and-validation genes} include program analyses, simulation, diagnostics, mitigation, and benchmark oracles. LintQ and MorphQ validate program and transformation properties, whereas MQT Bench and Bugs4Q provide cross-level artifacts and reproducible evidence~\cite{paltenghi2024lintq,paltenghi2023morphq,quetschlich2023mqtbench,zhao2023bugs4q}. Although these systems belong to the same family, they validate different contract subjects.

% \textbf{Empirical Grounding.}
% We ground the taxonomy in documented features of 14 public systems. The
% % examples above illustrate representative mappings, while the complete
% feature-to-gene assignments are provided in a machine-checkable evidence
% catalog. The catalog classifies individual responsibilities rather than entire
% repositories because one system may contain genes from several strata and
% families.

\textbf{Scope and Relative Completeness.} The taxonomy covers responsibilities whose correctness depends on a quantum-specific state, operation, resource, capability, timing condition, noise model, or statistical obligation. Purely classical functionality remains outside this scope unless it affects such a contract.

Let $\mathcal{Q}$ denote the scoped responsibilities, $\mathcal{L}$ the software strata, and $\mathcal{F}$ the responsibility families. We define set-valued typings $\lambda:\mathcal{Q}\rightarrow 2^{\mathcal{L}},\tau:\mathcal{Q}\rightarrow 2^{\mathcal{F}}$, and the uncovered set $U=\{q\in\mathcal{Q}\mid \lambda(q)=\emptyset \lor \tau(q)=\emptyset\}$. A responsibility may receive multiple labels, provided that each denotes a distinct contract subject. We claim only relative completeness for the current catalog: new elements of $U$ may reveal missing categories or refine existing boundaries.

\begin{table*}[t] \caption{QGenome responsibility taxonomy. Each gene is typed by its software stratum.} \label{tab:taxonomy} \centering \scriptsize \setlength{\tabcolsep}{4pt} \renewcommand{\arraystretch}{1.02} \begin{tabularx}{\textwidth}{@{}p{1.45cm}p{2.05cm}X X@{}} \toprule \textbf{Group} & \textbf{Family} & \textbf{Semantic responsibility} & \textbf{Representative contract focus}\\ \midrule Semantic & State & Establish or encode a quantum state & Bell/GHZ preparation, ansatz initialization, state relation\\ Semantic & Transformation & Change quantum information coherently & QFT, oracle, compute--uncompute, pre/post relation\\ Semantic & Interaction & Create or consume quantum relations & Controlled operations, entanglement, resource roles\\ Semantic & Observation & Produce interpretable evidence & Basis measurement, syndrome processing, bit mapping\\ \midrule Coordination & Resource lifecycle & Govern resource phases & Allocation, reset, reuse, cleanliness, ownership\\ Coordination & Orchestration & Order quantum and classical work & Feedforward, hybrid loops, dependency, latency\\ \midrule Realization & Translation & Lower representations while preserving meaning & Decomposition, optimization, semantic refinement\\ Realization & Mapping \& timing & Bind logical space and time to a target & Placement, routing, scheduling, map consistency\\ Realization & Control \& calibration & Bind operations to physical signals & Pulse and readout synthesis, calibration freshness\\ \midrule Assurance & Protection \& validation & Establish or preserve software claims & QEC, mitigation, analysis, simulation, oracles\\ \bottomrule \end{tabularx} \end{table*}

\section{Genome Lifecycle Construction}

\textbf{Construction and Composition.} A quantum software genome instantiates the taxonomy by connecting typed gene cards into executable pathways. Genome construction identifies semantic responsibilities, assigns their strata and families, binds concrete expressions, and records dependency and refinement relations together with versioned evidence. Different software classes naturally exhibit different profiles: applications emphasize computational semantics, compilers emphasize translation and mapping, runtimes emphasize coordination, and controllers and analysis tools emphasize control and assurance. These profiles are not fixed labels, since one system may own several gene types and one gene may have multiple backend-specific expressions. Static analysis, metamorphic testing, simulation, execution, and calibration can provide evidence at different strata~\cite{paltenghi2023morphq,paltenghi2024lintq}.

\textbf{Contextual Validity.} A genome is valid only under a declared execution context. Its genes must satisfy local contracts, compose consistently, preserve higher-level responsibilities across refinement, and rely on applicable evidence: $\textit{Valid}(\Gamma,K)=
\textit{Local}(\Gamma,K)\land
\textit{Compose}(\Gamma)\land
\textit{Refine}(\Gamma,K)\land
\textit{Fresh}(V,K)$. A successful execution alone is insufficient, as it may conceal mapping errors, untested branches, or stale calibration evidence.

\textbf{Genome Evolution.} A genome evolves through changes to its responsibilities, pathways, expressions, contexts, or evidence. \textit{Expression variation} changes an implementation while preserving its contract, such as a revised gate decomposition or pulse schedule. \textit{Pathway recombination} changes gene composition, as when qubit reuse introduces measurement and reset responsibilities~\cite{decross2023reuse}. \textit{Contract mutation} instead changes the promised semantics, such as replacing an exact transformation with an approximate one. These distinctions separate contract-preserving implementation changes from semantic changes or violations across genes.

Environmental adaptation re-expresses a stable responsibility under a new execution context $K'$. Retargeting is valid only when $K'\models A$, $(X',K')\models G$, and the evidence $V'$ adequately supports these claims. Depending on the contract $\Phi$, the correspondence between the original and adapted expressions may be exact, approximate, or statistical. This distinction is particularly important for hardware-facing systems such as Pulser and Qibolab, where device capabilities and configurations affect realization validity~\cite{silverio2022pulser,efthymiou2024qibolab}. Phenotypic measures such as fidelity, latency, success probability, and resource cost may guide the selection among valid adaptations, but do not determine gene identity.

\section{Capability Expression and Engineering Uses}

\textbf{From Validity to Capability.} QGenome connects cross-stratum validity to observable software capabilities. A \textit{phenotype} is an observation produced by a genome under a declared context, whereas a \textit{capability} is a set of supported phenotypes with explicit operating conditions. This distinction prevents a successful execution from being generalized into an unsupported capability claim.

Phenotypes vary across software strata. Applications expose output quality and protocol success, compilers expose semantic preservation and resource overhead, runtimes expose execution and feedback behavior, and control software exposes signal validity and calibration quality. Assurance tools further characterize detected violations and validation coverage.

A capability claim is justified only when the relevant genome pathway is valid under the declared context and the collected observations satisfy its acceptance criteria with sufficient evidence. Capability assessment must therefore distinguish the validity of the pathway, the conditions under which it is realized, and the evidence supporting its observed behavior.

\textbf{Engineering Uses.} By making responsibilities, contexts, and evidence explicit, QGenome supports several engineering tasks. \textit{Reuse} transfers gene cards and pathways together with their assumptions, so incompatible contexts trigger adaptation rather than silent transplantation. \textit{Testing} measures coverage over responsibilities and refinement edges instead of only statements or gates. Existing resources such as MorphQ, MQT Bench, and Bugs4Q can provide evidence for these checks~\cite{paltenghi2023morphq,quetschlich2023mqtbench,zhao2023bugs4q}. \textit{Repair} can target the earliest violated contract, while change-impact analysis follows dependency, refinement, and evidence-freshness relations.

Gene cards can also serve as typed planning artifacts for autonomous engineering agents. An implementation agent proposes an expression, a backend-aware agent checks its context and resource assumptions, and a verification agent supplies supporting evidence. Existing verifier-in-the-loop and closed-loop QPU systems show that such executable feedback is increasingly available~\cite{yu2026vista,cao2025agents}. QGenome does not make an agent trustworthy by itself. Instead, it specifies the contract, context, and evidence that must be established before an agent-generated action is accepted.
\section{Future Research Directions}

\textbf{QGene Atlas and Benchmark.}
A first direction is to build a versioned atlas of quantum software genes across applications, compilers, runtimes, control systems, simulators, and assurance tools. The atlas should record gene cards, pathway relations, concrete expressions, contexts, and evidence rather than treating repositories as indivisible samples. It can support tasks such as gene-boundary detection, family classification, cross-stratum linking, and contract extraction. Annotation agreement and unresolved cases should guide future revisions of the taxonomy.

\textbf{Automated Gene Mining and Lineage Recovery.}
Future work can recover gene cards and their evolution from software artifacts. Static analysis can identify resource roles and transformation boundaries, while execution traces reveal runtime dependencies and hardware interactions. IR provenance and version history can link source responsibilities to compiled and deployed expressions. A practical interchange format should capture expressions, refinement relations, evidence validity, and phenotype observations.

\textbf{Gene-Aware Regression Testing and Repair.}
QGenome enables cross-layer regression analysis beyond circuit syntax. Compiler or calibration changes may violate contracts without changing the program. Future systems can mutate gene cards and refinement edges, generate contract-oriented tests, and localize failures to violated responsibilities. Versioned compiler, runtime, and calibration benchmarks can further evaluate gene-aware fault localization and contract-guided repair.

\textbf{Evidence-Grounded Quantum Software Agents.}
Quantum software agents should produce validated genome updates, not just circuits or configurations. Each update should specify affected responsibilities, required context, proposed expressions, and supporting evidence. This supports backend-aware mapping, testing, workflow adaptation, repair, and retargeting. Evaluation should consider contract coverage, unsupported assumptions, evidence freshness, cross-layer consistency, and adaptation to environmental changes alongside task success.

\section{Conclusion}
In this paper, we introduced \textit{QGenome}, a semantic foundation for understanding quantum software beyond gates, circuits, and repository boundaries. We define quantum software genes as typed, reusable responsibilities with explicit contracts, contexts, expressions, and evidence, and organize them across five software strata and ten responsibility families. QGenome provides a common vocabulary for tracing how quantum-dependent responsibilities are realized, composed, and preserved across the stack, enabling future gene-aware analysis, testing, repair, evolution, and autonomous quantum software engineering.

% \section*{Acknowledgement}

% We thank all reviewers for their insightful comments and suggestions.
% This research is supported by Huawei, the National Research Foundation, Singapore, the Cyber Security Agency under its National Cybersecurity R\&D Programme (NCRP25-P04-TAICeN), and CyberSG R\&D Programme Office under its Translation and Innovation Grant (CRPO-GC5-SMU-012). Any opinions, findings and conclusions or recommendations expressed in this material are those of the author(s) and do not reflect the views of Huawei, the National Research Foundation, Singapore, the Cyber Security Agency of Singapore, and CyberSG R\&D Programme Office.

\bibliographystyle{ACM-Reference-Format}
\bibliography{ref}

\end{document}